\documentclass[manuscript,screen,table]{acmart}

\begin{document}

\title[How Should We Support Designing Privacy-Friendly Apps for Children?]{How Should We Support Designing Privacy-Friendly Apps for Children? Using a Research through Design Process to Understand Developers' Needs and Challenges}

\author{Anirudh Ekambaranathan}
\email{anirudh.ekam@cs.ox.ac.uk}
\author{Jun Zhao}
\email{jun.zhao@cs.ox.ac.uk}
\author{Max Van Kleek}
\email{max.van.kleek@cs.ox.ac.uk}
\affiliation{%
  \institution{Department of Computer Science \\ University of Oxford}
  \city{Oxford}
  \country{United Kingdom}
}

\renewcommand{\shortauthors}{Ekambaranathan, et al.}

\begin{abstract}
Mobile apps used by children often make use of harmful techniques, such as data tracking and targeted advertising. Previous research has suggested that developers face several systemic challenges in designing apps that prioritise children's best interests. To understand how developers can be better supported, we used a Research through Design (RtD) method to explore what the future of privacy-friendly app development could look like. We performed an elicitation study with 20 children's app developers to understand their needs and requirements. We found a number of specific technical requirements from the participants about how they would like to be supported, such as having actionable transnational design guidelines and easy-to-use development libraries. However, participants were reluctant to adopt these design ideas in their development practices due to perceived financial risks associated with increased privacy in apps. To overcome this critical gap, participants formulated socio-technical requirements \textcolor{black}{that} extend to other stakeholders in the mobile industry, including parents and marketplaces. Our findings provide important immediate and long-term design opportunities for the HCI community, and indicate that support for changing app developers' practices must be designed in the context of their relationship with other stakeholders.
\end{abstract}



\maketitle

\section{Introduction}
The pace at which technological advancements have been made in the past few decades has shifted many children's activities from the physical to the digital. Children are spending more time online through mobile devices now than ever before \cite{ofcom2018children, livingstone2017children, ofcom2018children}. A recent report shows that nearly all children aged 3 - 17 in the UK have been spending time online, primarily through mobile devices such as smartphones (72\%) and tablets (69\%) \cite{britain2022children}, and over 60\% of children under the age of 13 have a social media profile \cite{britain2022children}. Unsurprisingly, the industry for developing, distributing, and monetising services and applications aimed at children is thriving. Mobile marketplaces, such as the Google Play Store and the Apple App Store, all offer specific categories of apps aimed at children \cite{google_expert, apple_arcade, google_play_pass}.

The increased use of mobile applications by children does not come without risks and harms \cite{livingstone2017children}. Data privacy is one such risk, a large portion of which can be attributed to the use of third-party development libraries, which have now become an essential part of mobile app development \cite{balebako2014privacy}. App developers often opt for these libraries for the ease of development, but more often for data monetisation. These libraries are often embedded with data trackers to collect sensitive user data \cite{book2013longitudinal, lin2013understanding, reyes2018won}, including location information, and share processed data with data brokers for analytics purposes. This issue is particularly concerning for children~\cite{binns2018third} as research has shown that ``family apps" are often associated with more trackers than many other app genres \cite{livingstone2017children}. Furthermore, children's understanding of such digital risks is less developed than that of adults \cite{livingstone2017children, lapenta2015youth, emanuel2014exploring, zhao2019make}, raising a critical need \textcolor{black}{to} explore how we can better support children's privacy rights and autonomy \cite{peter2011adolescents, raynes2014gaming, pradeep2016virtual, balleys2017being}.

Over the past few years, concerns have been expressed by regulatory bodies, human rights organisations, and industry stakeholders that children's privacy and the commercial use of their data \textcolor{black}{are} at risk \cite{lupton2017datafied}. A notable direction in which efforts have been made to improve the state of privacy for children is by rethinking the role of developers and service providers. For example, the UK Information Commissioner’s Office (ICO) introduced the statutory Age-Appropriate Design Code (AADC) \cite{ico_2020} in 2021, which requires services and applications aimed at children to consider data protection as a key element as part of their design. However, enforcing compliance is not without challenge. For this study, we broadly refer to being \textit{``privacy-friendly''} as the design principles that minimise the amount of implicit personal data collection and sharing in apps through the use of data trackers embedded in third-party app development libraries and advertising modules. We are interested to examine how app developers for children may need to be better supported for building privacy-friendly apps, given the increased legislation developments in the UK.

In recent years, the human-computer interaction (HCI) community has seen growth in research involving app developers. Despite this progress, much of the research has primarily focused on understanding app developers' perceptions and practices \cite{li2021developers, tahaei2021privacy, ekambaranathan2020understanding, mhaidli2019we}. For example, it has been shown that developers often choose SDKs and third-party libraries based on their popularity, rather than considering their potential impacts on user privacy and security \cite{mhaidli2019we, ekambaranathan2021money}. Additionally, app developers frequently maintain default configurations for these tools \cite{tahaei2021developers}, which may not adequately protect users' data. Developers have also been found \textcolor{black}{to} often lack practical support, for example through guidelines and SDKs, for their implementation of privacy-friendly features, and face systemic monetary constraints placed on them by marketplaces, leading to them having to make a trade-off between sustaining their business and protecting children's privacy \cite{ekambaranathan2020understanding, ekambaranathan2021money}. While all these studies point to the need for usable tools to help app developers, research in this direction is still limited, and few efforts have been made to support them through concrete tools. 

Therefore, in this study we \textcolor{black}{aim} to understand how these practical and structural challenges can be overcome. We make use of \textcolor{black}{the} Research through Design (RtD) approach \cite{zimmerman2007research} to explore what future tools could look like \textcolor{black}{that would} support developers in creating privacy-friendly apps for children. We were guided by the following research questions:

\begin{enumerate}
    \item How do app developers want specific development tools to enable better privacy in children's apps to look like?
    \item What requirements do app developers foresee for tools to support their development of future privacy-friendly apps for children?
    \item How do app developers envision overcoming systemic barriers, imposed by major marketplaces, to creating privacy-friendly apps?
\end{enumerate}

The socio-technical nature and complexity of the challenges we \textcolor{black}{aim} to address mean that potentially disruptive ideas are needed to overcome the systems currently put in place. For this reason, we made use of an RtD method, which is a co-design methodology that allows researchers to design and discuss ideas \textcolor{black}{that} may seem futuristic, disruptive and even technically infeasible to realise \cite{stappers2017research}. Using a set of design ideas generated by expert designers, we were able to engage with app developers more effectively, elicit their technological requirements for integrating privacy considerations as a part of the development process. Previous research has shown that RtD is particularly effective for identifying users' latent needs in support of `wicked problems', which can be challenging to design for~\cite{zimmerman2007research,rittel1973dilemmas}. To this end, we performed an ideation workshop with expert designers in which we generated 12 speculative designs aimed at supporting developers in creating privacy-friendly apps. We presented and discussed these designs with 20 children's app developers through semi-structured interviews to elicit their reactions and understand what their requirements are when envisioning the future of privacy-friendly app development.

Our findings confirmed the technical and systemic barriers identified in previous research and identified \textcolor{black}{specific} thoughts and requirements from our participants regarding how practical tools and guidance should look like. Furthermore, although our results showed that participants were wary \textcolor{black}{of} addressing systemic challenges through some of the more disruptive ideas we proposed (e.g. through alternative marketplaces), they formulated critical socio-technical barriers that prevent current practice changes and indicated the need for a multi-stakeholder multi-disciplinary approach \textcolor{black}{to} understanding the interplay of incentives and behaviours of multiple stakeholders, including parents, children, and marketplaces. Our findings provide important implications and opportunities for design, the potential for \textcolor{black}{changing} power structures in the mobile ecosystem, and key directions for concretely supporting developers.

\section{Related work}

In this section, we describe related work on privacy practices in mobile apps for children and the role of developers herein, and provide background on the Research through Design method.

\subsection{Children's privacy in the mobile ecosystem} 
Despite the fact that children nowadays grow up in a digital environment, they are considered to be particularly vulnerable to the risks of data collection. The majority of them have a lesser understanding of privacy related contexts \textcolor{black}{and its associated digital risks }  \cite{livingstone2017children} than most adults \cite{zhao2019make, kumar2017no}. They often struggle to fully understand how and why their data is collected by third parties \cite{lapenta2015youth, emanuel2014exploring, acker2018youth}, how it is processed \cite{bowler2017lives}, or how it can be used in the future \cite{murumaa2015drawing,bowler2017lives,pangrazio2018s}. Children have grown to accept targeted advertising and analytics as part of everyday life \cite{lapenta2015youth}, without being able to change their behaviours to affect this \cite{lapenta2015youth, pangrazio2018s}. They feel forced to comply with privacy conditions set out by services \cite{lapenta2015youth} and find it difficult to understand them due to their length and legal jargon \cite{best2017growing}. However, it has been shown that children do value their privacy, so that they can enjoy their online experiences \cite{kumar2017no, silva2017privacy}. They can construct analogies between the digital and real world for privacy scenarios, such as equating concepts of ``hiding secrets'' to ``keeping things to yourself'' \cite{zhang2016nosy}.  While children do want to preserve their privacy, their understanding of the risks associated with long term accumulation of data is still underdeveloped \cite{pangrazio2017my, wang2017understanding}.

One significant factor contributing to \textcolor{black}{children's loss of online privacy} is the use of third-party libraries in app development, which often involves the collection of data for targeting purposes. While these libraries can provide benefits such as simplifying development, improving security, and adding additional functionality to apps \cite{enisa2018}, they have also been found to access location permissions, track call logs, browser history, and contact information for the purpose of targeted advertising \cite{grace2012unsafe}, and this data is often sold to advertisers through data brokers. Data trackers are found to be particularly prevalent in apps in the ``family'' category of the Google Play Store~\cite{binns2018third}, with these apps associated with the second highest number of data trackers, indicating that children's data is at risk of being collected and potentially misused \cite{binns2018third}.

The prevalence of data tracking features and the loss of privacy for children can lead to concrete harms which are often overlooked, such as identity theft and fraud \cite{coughlan2018sharenting}, the \textcolor{black}{normalisation of a culture where data surveillance is commonplace} \cite{kobie2016surveillance}, and long term risks to children's reputation and opportunities as they grow older \cite{longfield2018knows}. As a result, the protection of children's privacy is considered particularly important. The datafication and tracking of children's data have become the focus of several recent regulatory developments, such as UK's Age Appropriate Design Code~\cite{ico}, Online Safety Bill development~\cite{onlineharm}, and the ongoing development of the California Age Appropriate Design Code \cite{ccaadc}.

\subsection{Developer practices in the mobile ecosystem}
Initiatives to control the direction of mobile technological developments have primarily been aimed at \textcolor{black}{major stakeholders in the industry}, such as Google, Apple, and Facebook. The role of service providers and app creators, whilst having gained more traction over the past few years, has remained largely understudied in \textcolor{black}{the context} of digital harm and \textcolor{black}{in} the HCI community. In this research, we refer to that actors involved in the development of apps, such as UX designers and engineers, broadly as `developers'. In the HCI community we are starting to develop a better understanding of design practices used by developers in apps as well as challenges developers face preventing them from creating privacy-friendly apps~\cite{assal2019think,balebako2014privacy,mhaidli2019we}, which may have directly contributed to the app landscape we see at the moment. 

One of the reasons why data tracking through third-party libraries is prominent in apps is that developers rely on targeted advertising for generating revenue \cite{leontiadis2012don, acquisti2016economics, interactive2015iab}, which often depends on the third-party libraries to collect data. The ad networks that developers integrate in their apps often use low privacy settings by default \cite{tahaei2021developers}, which developers often keep unchanged \cite{mhaidli2019we}, meaning that these networks collect more data than is necessary. In many cases, ad networks also have specific configurations for children's apps, which are rarely adopted by the developers. \textcolor{black}{Furthermore}, it has also been shown that developers can also be exposed to dark patterns when working with ad network APIs, nudging them into making choices detrimental to the privacy of end users \cite{tahaei2021developers}. Research with developers has shown that developers can be made more aware of this by making the privacy consequences apparent in the API interface \cite{tahaei2021developers}. Finally, the complex legal language is another contributing factor making it challenging for developers to understand and translate privacy requirements to technical features\cite{tahaei2021privacy, bednar2019engineering}. Research has shown that developers often \textcolor{black}{find} it difficult to navigate the various permissions on apps and write privacy policies for their apps, \textcolor{black}{which} is required by the marketplaces \cite{tahaei2021privacy, li2021developers}. 

We have broadly identified two types of challenges faced by developers. The first set of challenges are practical and technical in nature, and can be tackled by creating appropriate tooling and putting in place the necessary support systems. For example, recent work has shown that developers often choose SDKs and third-party libraries based on their popularity, rather than considering their potential impacts on user privacy and security \cite{mhaidli2019we, ekambaranathan2021money}; they also frequently maintain default privacy configurations for these tools \cite{tahaei2021developers}, which may not adequately protect users' data. \textcolor{black}{Similarly, developers find} that guidelines for designing for children are not always accessible and may conflict with data protection standards put forth by different organisations~\cite{ ekambaranathan2021money, ekambaranathan2020understanding}. These complex guidelines raise particular challenges for app developers who need to reduce the cost of committing to the building of privacy-friendly apps. Similarly, it has been shown that developers do not always fully understand the data collection behaviours of third-party libraries and APIs~\cite{balebako2014privacy, ekambaranathan2021money}, making them widespread in their development practices, affecting children's data privacy, as well as the level of transparency that they can provide.

Secondly, previous research \textcolor{black}{has} indicated that developers find themselves forced to comply with the practices encouraged or set forth by major marketplaces and corporations. For example, they feel compelled to employ revenue models that are often \textcolor{black}{based on methods using} targeted data analytics, such as advertising, primarily because such methods are conveniently offered by major marketplaces \cite{mhaidli2019we}. Moreover, they find that privacy-invasive apps, which are often offered for free, will always have a competitive advantage over more privacy-friendly and ethical apps which are offered for a premium \cite{ekambaranathan2021money}. These challenges speak to the structural nature of the current ecosystem, \textcolor{black}{as opposed to} focusing on individual tools and technologies, it requires a paradigm shift in the way apps are created, distributed and monetised, through a series of systemic changes. 

While plenty of tools are available to support secure app/software development, developers have \textcolor{black}{far fewer choices and less support} for building apps which do not extensively make use of third-party data trackers. We broadly adopt the term \textit{privacy-friendly} apps to refer to the set of apps which minimise the amount of personal data collection and sharing, for example through data trackers embedded in third-party libraries and advertising modules. As shown by our above discussions, previous research has primarily been focused on understanding developer behaviours \cite{ekambaranathan2020understanding, tahaei2021privacy}, rather than looking to supporting them through concrete tools. Thus, in this study, we wish to address the challenges preventing privacy-friendly development identified in previous research \cite{ekambaranathan2021money}, by exploring a range of design approaches for creating new tools to support privacy-friendly app development, and identifying future research directions. 

\subsection{Research through Design}
The term `Research through Art and Design' was first introduced by Frayling~\cite{frayling1993research}, discussing ways in which \textcolor{black}{conducting} research could be of interest to the design community. It was not intended to be aimed at interaction design specifically~\cite{stappers2017research}. \textcolor{black}{Most of the current academic literature on} `Research through Design' (RtD) stems from the HCI community. RtD was formalised in the HCI community by Zimmerman in the late 2000s \cite{zimmerman2007research, zimmerman2010analysis}, describing it as a ``research approach that employs methods and processes from design practice as a legitimate method of inquiry''. RtD is meant to be a method of generating knowledge through a design process, and is not intended to necessarily immediately produce a commercial product. 

The theoretical literature on RtD is still in its formative stages, and is only recently gaining traction in the HCI community. As a result, there is no one agreed upon definition of what RtD is, or how it should be conducted, or how knowledge through an RtD method should be generated. However, most practitioners agree that the process of doing design can constitute research and can lead to the generation of knowledge \cite{lunenfeld2003design}: ``The designing act of creating prototypes is in itself a potential generator of knowledge (if only its insights do not disappear into the prototype, but are fed back into the disciplinary and cross-disciplinary platforms that can fit these insights into the growth of theory).''~\cite{stappers2012doing}. Doing RtD generally involves the development of a prototype or design that shows a product which interacts with people in a way that was not possible before. This opens a discussion about how the future design space in that area could and should look like, focusing on identifying the \textit{latent needs} of stakeholders.

RtD has been particularly useful for enabling researchers to identify opportunities for new technologies~\cite{zimmerman2007research} by capturing a future form enabled with concrete and specific design artefacts. Previous studies have shown that these concrete embodiments have been effective for researchers and designers to explore the design space with the target stakeholders~\cite{zimmerman2017speed}, such as integrating smart home devices at home settings~\cite{davidoff2007rapidly} or introducing robotic care to the elderly without intimidating them~\cite{forlizzi2005sensechair}.

For many researchers, RtD is regarded as an effective methodology for examining ``wicked problems'' through exploring design artefacts that are intended to transform the world from its current state to a preferred future state \cite{zimmerman2007research}. Wicked problems generally refer to problems that cannot be accurately modelled by the reductionist approaches of science and engineering due to conflicts between stakeholders \cite{rittel1973dilemmas}. We can regard the problem of designing privacy-friendly apps as a wicked problem, because structural and systemic features of the app economy are actively discouraging developers from creating privacy-friendly apps. \textcolor{black}{Using RtD we can propose provocative and disruptive ideas to developers to explore how the future design space for creating privacy-friendly apps should look like \cite{stappers2017research}. }
\section{Methods}
In this study, we sought to understand how we can best support app developers in building privacy-friendly apps and to uncover their latent needs, perspectives, and values \textcolor{black}{that} need to be embedded in designing future tools and support systems. We made use of an RtD method, an alternative to other co-design methodologies, with its particular focus on gaining knowledge and research through the process of doing design. An RtD method allows us to start \textcolor{black}{by} creating an initial design space by involving design experts who can formulate potentially disruptive ideas \cite{stappers2017research}. The aim of this is to transgress what is currently possible within the limits of the current app ecosystem, \textcolor{black}{thereby provoking developers to think about how they envision the future of privacy-friendly app development and how they would like to be supported.} This initial design space may provide a more useful starting point for particularly exploring `wicked problems' that are challenging to capture and require creativity to imagine how the future could appear. As there is no strict definition of the experiments associated with RtD, we strengthened the robustness of our findings by drawing on existing HCI methods for our design activities and elicitation study. We made use of a three-phase process \textcolor{black}{that were} based on previous research using RtD \cite{chen2021happens}:

\begin{itemize}
    \item First, we ran an \textit{Ideation Workshop} with experienced HCI designers to explore a range of initial design ideas for supporting the creation of privacy-friendly apps, which we solidified as \textit{Speculative Sketches}~\cite{desjardins2020iot}. These are not intended to be perfect, but serve to quickly capture design ideas.
    \item Second, we converted the sketches into more intuitive and understandable illustrations, and included more detailed documentation of the design features for each idea, which we captured in a \textit{Design Workbook}\cite{blythe2018imaginary, gaver2011making, wong2017eliciting}. The design workbook is intended to work in a standalone format, where participants can engage with it and ideally understand the design concepts without needing intervention, and provide feedback through sketches and comments. 
    \item  Lastly, we conducted semi-structured interviews using a \textit{Speed Dating} \cite{davidoff2007rapidly, zimmerman2017speed} approach, where participants were presented with the ideas from our design workbook one by one. They then expressed their thoughts, shared concerns and needs \textcolor{black}{related to} the designs we proposed, and voted on their favourite design ideas. By using a speed dating approach, and being presented with a range of options, we aim to identify qualities participants are looking for in technologies to change the status quo \cite{zimmerman2017speed}.
\end{itemize}

\subsection{Ideation workshop}
The key to a successful application of the RtD method is to ensure that it creates an integration between theoretical knowledge and technical engineering of the knowledge. Thus, researchers are expected to have explored the knowledge and design space \textit{before} attempting to \textcolor{black}{create a final product}. They are expected to go through an active process of ``ideating, iterating, and critiquing potential solutions'', to continually reframe the problem in order to get closer to ``a concrete problem framing and articulation of the preferred state, and a series of artefacts—models, prototypes, products, and documentation of the design process''~\cite{zimmerman2007research}. 

For these purposes, we made use of an ideation workshop~\cite{zimmerman2007research,chen2021happens} in our study. Ideation workshops are typically short workshops that involve experienced designers from a related application domain to brainstorm and refine an initial design space. We organised a two-hour ideation and brainstorming workshop, \textcolor{black}{which was initially planned as an in-person event, but due to COVID-19 limitations, it was held virtually}. Our workshop was based on the Design Sprint method~\cite{knapp2016sprint}, which is used to rapidly prototype and evaluate ideas, and includes three stages: identifying `How Might We (HMW)' questions, lightning demos (seeking for existing solutions in parallel domains), and solution sketching. 

In selecting participants for the ideation workshop, we followed the Design Sprint recommendation to limit the workshop to ``seven or fewer'' people \cite{knapp2016sprint}. We were specifically looking for participants who had prior \textcolor{black}{experience} in app development for children, knowledge about data protection and privacy for children, and optionally any industry experience. By maintaining these criteria, we ensured that participants could relate to the design challenges at hand and draw from their own experiences and knowledge to generate meaningful and expert-led design ideas. Using these criteria, we invited three final-year doctoral HCI researchers, a senior research associate from the same research group, as well as the first two authors. All participants have extensive \textcolor{black}{experience} in app development and user-facing designs for children. In addition, the first author, two of the researchers, and the research associate have industry experience in software and app development. All participants have experience in HCI research and are all very well-versed in matters relating to privacy for children in apps. 

At the start of the workshop, we contextualised the problem space by explaining to the participants (researchers and designers) the specific social-technical challenges that app developers are facing that were identified in the previous work~\cite{ekambaranathan2021money}, which include the following: 
\begin{itemize}
\item \textit{The need for accessible privacy best practices guidance}: What can we support developers to translate existing policy guidance and regulations on privacy into practice? What should good and useful design guidelines look like for developers?
\item \textit{The need for designing tools to enforce best privacy practices}: What kind of technical toolkits are needed by developers for building privacy-friendly apps? 
\item \textit{The need for addressing systemic issues in marketplaces}: How would developers like to be supported in a market undervaluing responsible development for users' best interests and an ecosystem lacking reward and recognition for such practices?
\end{itemize}

\textit{HMWs.} In the first stage of the workshop, all the participants were asked to identify problem areas for these challenges by individually writing down ``How Might We'' (HMW) questions (e.g., ``How might we make data protection regulations more accessible to developers?''). This was followed by participants reading through their HMWs to the group so that each participant could vote on the three most interesting ones they wanted to tackle as a group. The three top-voted HMWs were used as a starting point to brainstorm solutions in the next stage. 

\textit{Lightning demos.} In the second stage of the workshop, participants were asked to go online and seek existing solutions to the HMWs from similar problem domains to inspire ideas for solution sketching. These were then presented in short demos for all participants so that they could become familiar with them. It was \textcolor{black}{essential} that all participants had prior knowledge and experience with designing mechanisms to promote better user privacy awareness or translating data protection guidelines \textcolor{black}{into} more actionable design candidates, to ensure the effectiveness of this stage.

\textit{Solution sketching.} In the last stage, each participant was asked to reflect on the proposed design options and individually create 2-3 low-fidelity \textit{Speculative Sketches}. This was followed by a general discussion as a group reflecting on useful features and design rationales.

\subsection{Design workbook}\label{sec:book}
After the workshop, the first author went through all the sketches and categorised them according to the three \textcolor{black}{socio}-technical challenges presented in the ideation workshop. To ensure the collection of sketches as a whole would uniformly represent the research questions (see Section 1), we merged closely related concepts and discarded concepts which did not directly address the research question. For example, we discarded the initial sketch of a Privacy Guard concept, which was aimed \textcolor{black}{at blocking} network traffic to and from any advertising and tracking domains, because we felt this was aimed more at children rather than supporting developers.

We were left with 12 design concepts, which were redrawn to ensure uniformity. We added brief descriptions to highlight the core concepts and interaction features, and ended up with 12 concepts as our design workbook, illustrated in Figure~\ref{fig:book}. The full design workbook is available in the supplementary materials. 

 These designs address the three identified problems in the following ways:
\begin{itemize}
    \item \textbf{Supporting accessible development guidance.} Our first three ideas were aimed at reducing the complexity of guidance for privacy-friendly design \cite{ekambaranathan2021money}. The first design, a \textit{Dos and Don’ts Checklist}~\textbf{\texttt{[D1]}}, was aimed to concretise and simplify requirements set forth in common legal documents, such as GDPR and ICO's AADC. This way, complex principles are broken down into technical requirements, which developers have been shown to struggle with \cite{bednar2019engineering, tahaei2021privacy}. We also proposed a more formal \textit{training course}~\textbf{\texttt{[D2]}}, providing them with an opportunity of self-education and certification. This would allow developers to get a more holistic and broad understanding of important aspects in designing for children, including \textcolor{black}{reasoning} and motivations about why certain design features would be important. Lastly, \textit{Requirement Matrix}~\textbf{\texttt{[D3]}} was proposed to collate requirements from various marketplaces and legal frameworks. As apps are often launched into an international market, such a matrix can potentially enable developers to quickly assess their compliance with requirements from different marketplaces (e.g., Google and Apple) and different jurisdictions (e.g., US and EU).
    
    \item \textbf{Providing tools to support best practices.} This category of designs was aimed to support simplifying compliance assessment and proposing SDKs that can improve transparency and act in children's best interests. We proposed three methods to help in matters related to compliance. Firstly, we wanted to involve additional stakeholders (such as parents) in a crowd sourced mechanism to assess compliance with common data protection requirements, e.g., the AADC (\textit{Wisdom of the Crowds}~\textbf{\texttt{[D4]}}). Secondly, we proposed an automated method of doing this (\textit{Machine over Mind}~\textbf{\texttt{[D5]}}), and thirdly we proposed this compliance \textcolor{black}{being} carried out by an authoritative regulation/standardisation body (\textit{Universal compliance service}~\textbf{\texttt{[D6]}}). Furthermore, we also wanted to explore how developers may perceive the trade-off between carrying out additional development efforts in order to commit to transparency and children's best interests. We therefore proposed SDKs which developers could simply integrate into their apps without additional efforts, thereby removing any friction in this process. The \textit{Plug-and-play self-control tool}~\textbf{\texttt{[D7]}} adds a new interface to an existing app allowing children and parents to control screen time and change privacy settings. The \textit{Parental assistance}~\textbf{\texttt{[D8]}} tool allows developers to generate guidance for parents which subsequently can be accessed through the app.
    
    \item \textbf{Addressing systemic challenges.} In our final set of design ideas, we wanted to explore how we can overcome systemic challenges arising from marketplaces to incentivise developers to adopt privacy-friendly development principles. We looked at two approaches in achieving this. Firstly, we sought to help developers build recognition amongst consumers. Motivated by earlier design work to help users recognise data protection practices through `privacy labels'~\cite{kelley2009nutrition}, we proposed self-certified \textit{Public Pledges}~\textbf{\texttt{[D9]}}, awarded by developing technology according to certain standards, and a \textit{Badge of Honour}~\textbf{\texttt{[D10]}},  provided by a credible institution for upholding certain standards. Secondly, as targeted advertising is a commonly used method for developers to monetise their apps \cite{ekambaranathan2021money, mhaidli2019we, leontiadis2012don}, we looked at concretely providing alternative methods of financial support. Inspired by the open source innovation model and the decentralised data governance for promoting self-autonomy~\cite{stateofnation}, we also explored how developers perceive alternative marketplace models, such as \textit{Patreon and Crowdfunding}~\textbf{\texttt{[D11]}} for privacy-friendly apps or developing an \textit{Unlimited Arcade}~\textbf{\texttt{[D12]}} for promoting privacy-friendly and age-appropriate design approaches. 

\end{itemize}

\begin{figure}[h]
\begin{center}
\includegraphics[width=1.05\textwidth]{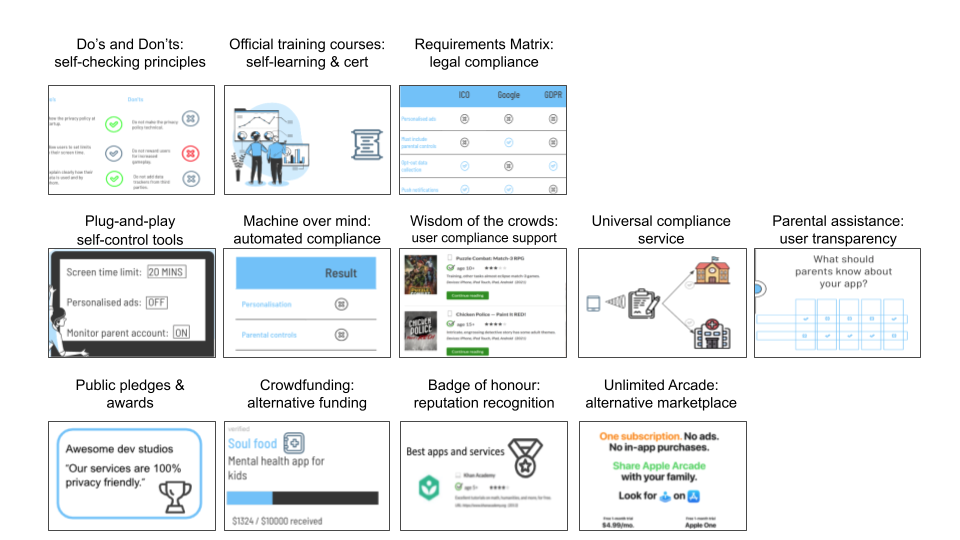}
\caption{An overview of the 12 concepts included in the design workbook. The text above each image icon provides a brief title of the design approach and a detailed description can be found in the supplementary materials.}
\label{fig:book}
\end{center}
\end{figure}

The design workbook was originally meant to be presented in a physical format, allowing participants to actively make notes and add ideas. However, as we conducted the workshops online, we transitioned to a digital format and presented the ideas as a slideshow through screen sharing. 

\subsection{Speed dating}
The speed dating design method~\cite{davidoff2007rapidly,zimmerman2017speed} draws on the analogy to a traditional social interaction that helps singles find a romantic partner, by providing rapid encounters with a series of potential romantic partners. The objective is to help participants to gain a better idea of what they \textcolor{black}{are looking} for in a partner and what they really want and need. Similarly, the speed dating design method places target stakeholders in a situation that enables them to sample a series of \textcolor{black}{possible} futures. After rapidly experiencing several possible futures, the study participants would then be guided to reflect \textcolor{black}{on} their implicit needs and barriers for adopting these possible futures. Because speed dating allows participants to reflect on concerns and needs which would not have arisen during fieldwork, it is effective to explore design opportunities in depth. The key for applying a speed dating design method is to put study participants \textcolor{black}{in} a situation that is familiar to them (such as recording TV programmes at home). Thus, during our study, when presenting each design idea to the study participants, we encouraged the app developer\textcolor{black}{s} to imagine how they may make use of the design in their actual app design and development process.

\subsubsection{Participants}
In this study, we specifically focused on app creators from the UK market. We also realise that the process of creating apps is complex and can involve multiple actors. The UK ICO broadly refers to this group of actors as developers, but acknowledges that this can include ``coders, UX designers and system engineers" \cite{ico_2020}, who are all responsible for upholding children's privacy in their development practices. We adopt a similar terminology in this study, and refer to app developers as any actor who may be involved in the process of creating an app.

\textcolor{black}{To recruit developers, we contacted development studios, product owners, and indie developers who had experience creating apps for children. We also included executives and directors, who have an important say in the app design process. We used search engines like Google and DuckDuckGo, and searched for keywords like "Kids App Developers UK" and "Game Studio UK" to find potential participants. We verified that companies were based in the UK through their contact and about pages. Participants were given a £15 Amazon gift voucher as compensation for participating. }

In total, we \textcolor{black}{recruited} 20 participants, with most participants being located in the UK. Participants not located in the UK still developed apps for the UK market. Two participants were female, the rest were all male. The participant characteristics are described in Table \ref{table:demographics}. Participants had varying roles in their organisations, with different levels of professional development experience in the sector, ranging from 15+ years to just starting out (<1 year). The roles of Director, Product Owner, Manager, and CTO, held by 14 of our participants, were included in our study because they had a significant influence over what the apps should look like and which principles they ought to adhere to. These participants would have direct/close involvement in ensuring the privacy-friendly requirements set out by the ICO and the marketplaces. The roles of Developer and UX Designer were held by the remaining 6 participants, and all except for one had experience in \textcolor{black}{indie app} development. One of the participants had just started their app development career, who was also included in the study as we felt that he could express which areas of privacy-friendly app development would need improvement for people just starting out.

\begin{table*}[t]
\vspace{1em}

\caption{\label{table:demographics}Table containing descriptive characteristics of the study participants. While all participants produced apps for the UK market, some were located in a different country. }
\begin{tabular}{lp{0.12\linewidth}lll}
\hline
\textbf{Participant} & \textbf{Country} & \textbf{Gender} & \textbf{Role} & \textbf{Years of experience}                          \\ \hline
P1            & UK              & M                         & Director                  & 16                                 \\ \hline
P2            & UK              & M                         & Operations director       & 15+                                 \\\hline
P3            & UK              & M                         & Head of product           & 10+                                 \\\hline
P4            & UK              & F                         & Head of production        & 9                                 \\\hline
P5            & UK              & M                         & Developer                 & < 1 (4 weeks)                                 \\\hline
P6            & UK              & M                         & Developer                 & 6                 \\\hline
P7            & UK              & M                         & Product designer          & 6                  \\\hline
P8            & UK              & M                         & Product owner             & 11                  \\\hline
P9            & UK              & F                         & Managing director         & 15+                           \\\hline
P10           & UK              & M                         & Technical director        & 20+                                \\\hline
P11           & UK              & M                         & UX Designer               & 3                                \\\hline
P12           & Germany         & M                         & Product owner              & 5+                                \\\hline
P13           & UK              & M                         & CTO                       & 10+                         \\\hline
P14           & UK              & M                         & Developer                 & 3           \\\hline
P15           & UK              & M                         & Developer                 & 5+   \\\hline
P16           & UK              & M                         & Product owner             & 10+                       \\\hline
P17           & UK              & M                         & Product owner             & 11                \\\hline
P18           & UK              & M                         & UX Designer               & 8                                                 \\\hline
P19           & Bulgaria        & M                         & Program manager           & 4                     \\\hline
P20           & Lithuania       & M                         & Product owner             & 10+                                      \\ \hline
\end{tabular}

\end{table*}

\subsubsection{Procedure and data collection}
We conducted 20 speed dating semi-structured interviews in the summer of 2021 using remote meeting software, such as Microsoft Teams or Zoom, depending on the participants’ preferences. Although this may not be the observational studies that we would intend in an ideal situation due to the disruption of the pandemic, we believe this closely resembled the app building situations as those were also carried \textcolor{black}{out} in a computer-based context. All discussions were audio recorded with the participants’ consent. Each session consisted of three parts. We started with a warm-up session in which we explained our research objectives and challenges. We asked about participants’ experiences in designing apps for children, what they think is important to keep in mind, and what they think challenges are for developers in privacy-friendly design. Then, the researcher shared their screen to display the design workbook and invited the app developers to picture themselves using the design ideas during their app designing, building, debugging, or publishing situation. 

With every concept, we gave participants some time to read the description and view the sketch, and allowed them to ask any questions for clarification. After participants described their initial reactions, we followed up with a series of questions to explore their sentiments, value alignments, and why participants liked or disliked the features. For example, ``How would you feel if you had to pay for the certificate?'' We explained to the participants that they could be honest and critical in their feedback, even if they felt negatively about a design idea, without adverse consequences. We concluded each interview with a summary slide \textcolor{black}{that} had all \textcolor{black}{the} concepts briefly listed. We asked participants which concepts they particularly liked and would like to see implemented in practice.

Our study has been approved by our Institutional Review Board. Participants signed consent forms and were informed in advance about their rights in the study. For example, they were informed that all data would be anonymised and that they could withdraw from the study before a specified date. 

\subsection{Data analysis}
Our reported results are mainly based on the discussions with the app developers during the interviews. All the collected audio recordings were transcribed and anonymised. Data was analysed using a grounded, thematic approach~\cite{braun2006using} to develop codes and themes related to developers' \textit{challenges, perceptions, requirements and values}. The thematic coding process started by dividing the transcriptions into two equal-sized sets. The first two authors independently analysed the first set of transcriptions to derive an initial set of codes. They then met to consolidate and reconcile codes into a common codebook. The first author then completed coding the remaining half of the transcriptions using this common codebook. During this process, we also created a mapping between the codes and each of the 12 design concepts \textcolor{black}{so} that we \textcolor{black}{could} gain an overview of how each option was perceived by the study participant.
 
In our analysis we did not specifically look out for differences between subgroups within our participants. While we found that participants indicated preferences for using certain design ideas within the capacity of their jobs, they often viewed and analysed these ideas from a viewpoint \textcolor{black}{that was} independent from their official job roles as possible.

Interviews lasted between 38 and 75 minutes. The total duration of the recordings is 1043 minutes and the average session length is 52 minutes (SD=10.2).

\begin{table}[t!]
\caption{\label{table:overview} Overview of participants' reactions: lime indicate a positive perception,red for a negative perception, magenta for an idea being both positively and negatively received and gray indicates a Neutral reaction.}
    \centering
    \small
    \tabcolsep=0.09cm
    \begin{tabular}{|l|p{0.023\linewidth}|p{0.023\linewidth}|p{0.023\linewidth}|p{0.023\linewidth}|p{0.023\linewidth}|p{0.023\linewidth}|p{0.023\linewidth}|p{0.023\linewidth}|p{0.023\linewidth}|p{0.023\linewidth}|p{0.023\linewidth}|p{0.023\linewidth}|p{0.023\linewidth}|p{0.023\linewidth}|p{0.023\linewidth}|p{0.023\linewidth}|p{0.023\linewidth}|p{0.023\linewidth}|p{0.023\linewidth}|p{0.023\linewidth}|}
    \hline
 \textbf{Participants} & P1 & P2 & P3 & P4 & P5 & P6 & P7 & P8 & P9 & P10 & P11 & P12 & P13 & P14 & P15 & P16 & P17 & P18 & P19 & P20 \\ \hline
        Dos and Don’ts Checklist~\textbf{\texttt{[D1]}}  & \cellcolor{lime} & \cellcolor{lime} & \cellcolor{lime} & \cellcolor{lime} & \cellcolor{lime} & \cellcolor{gray} & \cellcolor{lime} & \cellcolor{magenta} & \cellcolor{lime} & \cellcolor{lime} & \cellcolor{lime} & \cellcolor{lime} & \cellcolor{lime} & \cellcolor{lime} & \cellcolor{magenta} & \cellcolor{lime} & \cellcolor{gray} & \cellcolor{lime} & \cellcolor{lime} & \cellcolor{magenta} \\ \hline
        Courses: Making it official~\textbf{\texttt{[D2]}}  & \cellcolor{lime} & \cellcolor{red} & \cellcolor{lime} & \cellcolor{lime} & \cellcolor{magenta} & \cellcolor{lime} & \cellcolor{red} & \cellcolor{magenta} & \cellcolor{lime} & \cellcolor{red} & \cellcolor{gray} & \cellcolor{magenta} & \cellcolor{magenta} & \cellcolor{lime} & \cellcolor{gray} & \cellcolor{lime} & \cellcolor{lime} & \cellcolor{lime} & \cellcolor{lime} & \cellcolor{red} \\ \hline
        Requirements Matrix~\textbf{\texttt{[D3]}}  & \cellcolor{lime} & \cellcolor{lime} & \cellcolor{lime} & \cellcolor{lime} & \cellcolor{gray} & \cellcolor{lime} & \cellcolor{magenta} & \cellcolor{gray} & \cellcolor{lime} & \cellcolor{lime} & \cellcolor{lime} & \cellcolor{lime} & \cellcolor{lime} & \cellcolor{lime} & \cellcolor{lime} & \cellcolor{lime} & \cellcolor{lime} & \cellcolor{lime} & \cellcolor{lime} & \cellcolor{magenta} \\ \hline\hline
        Wisdom of the Crowds~\textbf{\texttt{[D4]}}  & \cellcolor{lime} & \cellcolor{magenta} & \cellcolor{magenta} & \cellcolor{red} & \cellcolor{magenta} & \cellcolor{magenta} & \cellcolor{red} & \cellcolor{magenta} & \cellcolor{magenta} & \cellcolor{red} & \cellcolor{red} & \cellcolor{lime} & \cellcolor{red} & \cellcolor{red} & \cellcolor{red} & \cellcolor{magenta} & \cellcolor{red} & \cellcolor{magenta} & \cellcolor{red} & \cellcolor{lime}  \\ \hline
        Machine over Mind~\textbf{\texttt{[D5]}}  & \cellcolor{lime} & \cellcolor{lime} & \cellcolor{lime} & \cellcolor{lime} & \cellcolor{lime} & \cellcolor{magenta} & \cellcolor{lime} & \cellcolor{red} & \cellcolor{lime} & \cellcolor{red} & \cellcolor{lime} & \cellcolor{gray} & \cellcolor{magenta} & \cellcolor{lime} & \cellcolor{lime} & \cellcolor{red} & \cellcolor{lime} & \cellcolor{lime} & \cellcolor{lime} & \cellcolor{lime}  \\ \hline
        Universal Compliance Service~\textbf{\texttt{[D6]}}  & \cellcolor{gray} & \cellcolor{gray} & \cellcolor{lime} & \cellcolor{lime} & \cellcolor{gray} & \cellcolor{lime} & \cellcolor{red} & \cellcolor{lime} & \cellcolor{gray} & \cellcolor{red} & \cellcolor{lime} & \cellcolor{lime} & \cellcolor{lime} & \cellcolor{lime} & \cellcolor{red} & \cellcolor{gray} & \cellcolor{red} & \cellcolor{lime} & \cellcolor{lime} & \cellcolor{red}  \\ \hline
        Plug-and-Play Self-Control~\textbf{\texttt{[D7]}}   & \cellcolor{magenta} & \cellcolor{lime} & \cellcolor{lime} & \cellcolor{magenta} & \cellcolor{lime} & \cellcolor{red} & \cellcolor{magenta} & \cellcolor{lime} & \cellcolor{magenta} & \cellcolor{lime} & \cellcolor{lime} & \cellcolor{gray} & \cellcolor{lime} & \cellcolor{magenta} & \cellcolor{magenta} & \cellcolor{magenta} & \cellcolor{gray} & \cellcolor{lime} & \cellcolor{magenta} & \cellcolor{red}  \\ \hline
        Parental Assistance~\textbf{\texttt{[D8]}}   & \cellcolor{lime} & \cellcolor{red} & \cellcolor{lime} & \cellcolor{magenta} & \cellcolor{gray} & \cellcolor{lime} & \cellcolor{lime} & \cellcolor{red} & \cellcolor{gray} & \cellcolor{lime} & \cellcolor{gray} & \cellcolor{red} & \cellcolor{magenta} & \cellcolor{magenta} & \cellcolor{gray} & \cellcolor{lime} & \cellcolor{red} & \cellcolor{red} & \cellcolor{magenta} & \cellcolor{lime}  \\ \hline\hline
        Public Pledges \& Awards~\textbf{\texttt{[D9]}}  & \cellcolor{magenta} & \cellcolor{lime} & \cellcolor{lime} & \cellcolor{lime} & \cellcolor{lime} & \cellcolor{lime} & \cellcolor{magenta} & \cellcolor{red} & \cellcolor{lime} & \cellcolor{lime} & \cellcolor{lime} & \cellcolor{lime} & \cellcolor{lime} & \cellcolor{lime} & \cellcolor{red} & \cellcolor{lime} & \cellcolor{red} & \cellcolor{lime} & \cellcolor{lime} & \cellcolor{lime}  \\ \hline
        Badge of Honour~\textbf{\texttt{[D10]}}  & \cellcolor{lime} & \cellcolor{lime} & \cellcolor{lime} & \cellcolor{lime} & \cellcolor{lime} & \cellcolor{lime} & \cellcolor{lime} & \cellcolor{red} & \cellcolor{lime} & \cellcolor{lime} & \cellcolor{lime} & \cellcolor{lime} & \cellcolor{lime} & \cellcolor{lime} & \cellcolor{red} & \cellcolor{magenta} & \cellcolor{magenta} & \cellcolor{lime} & \cellcolor{gray} & \cellcolor{lime}  \\ \hline
        Patreon and Crowdfunding~\textbf{\texttt{[D11]}} & \cellcolor{lime} & \cellcolor{lime} & \cellcolor{magenta} & \cellcolor{magenta} & \cellcolor{magenta} & \cellcolor{lime} & \cellcolor{magenta} & \cellcolor{red} & \cellcolor{magenta} & \cellcolor{magenta} & \cellcolor{magenta} & \cellcolor{lime} & \cellcolor{gray} & \cellcolor{magenta} & \cellcolor{lime} & \cellcolor{magenta} & \cellcolor{magenta} & \cellcolor{gray} & \cellcolor{red} & \cellcolor{lime}  \\ \hline
        Unlimited Arcade~\textbf{\texttt{[D12]}} & \cellcolor{lime} & \cellcolor{red} & \cellcolor{lime} & \cellcolor{lime} & \cellcolor{magenta} & \cellcolor{magenta} & \cellcolor{magenta} & \cellcolor{lime} & \cellcolor{lime} & \cellcolor{red} & \cellcolor{magenta} & \cellcolor{lime} & \cellcolor{gray} & \cellcolor{magenta} & \cellcolor{lime} & \cellcolor{gray} & \cellcolor{red} & \cellcolor{magenta} & \cellcolor{magenta} & \cellcolor{red} \\ \hline

    \end{tabular}
    \caption*{\colorbox{lime}{\strut Positive}\colorbox{red}{\strut Negative}\colorbox{magenta}{ \strut Mixed}\colorbox{gray}{ \strut Neutral}}
\end{table}

\section{Results}

Here we report requirements and concerns that participants reflected during their interaction with our design options. We start with an overview of participants reactions to our 12 design options, and then we describe the key themes which emerged from our analysis: (1) perceptions of guideline support; (2) requirements for development libraries; (3) concerns for compliance checking; (4) suggestions for alternative funding mechanisms; and (5) challenges for recognition of best practices. The requirements and barriers formulated by our participants are summarised in Table \ref{table:results}.

\subsection{Overview of participants' reactions}

\textcolor{black}{Table \ref{table:overview} summarises participant reactions to different design concepts, categorised as Positive, Neutral, Negative, or Mixed (i.e., both positive and negative). Participants provided clear feedback on their likes and dislikes. For example, P13 said the following about the \textit{Requirements Matrix}~\textbf{\texttt{[D3]}}: \textit{``This is excellent. This talks to a lot of the things that are missing from the checklist"}. All participants had a minimum of two positive reactions, and each idea was positively received by at least one participant.}

In general, most participants reacted positively \textcolor{black}{to ideas that help make guidelines more accessible}. They felt that \textcolor{black}{the} current guidelines and regulations are complex to understand, and that a simple breakdown would help during the development cycle. Specifically, the \textit{Dos and Don’ts Checklist}~\textbf{\texttt{[D1]}} and the \textit{Requirements Matrix}~\textbf{\texttt{[D3]}} were received well, because they are non-intrusive, i.e. they do not have to be technically enforced and there is no cost associated with owning or consulting such a resource. Similarly, participants also liked approaches which help with marketing and increase exposure to customers, such as the \textit{Public Pledges \& Awards}~\textbf{\texttt{[D9]}} or \textit{Badge of Honour}~\textbf{\texttt{[D10]}}. Such tools do not fundamentally affect apps and there is no deterring cost associated with using them. 

Participants were \textit{generally sceptical} about approaches which aimed to tackle structural problems. For example, through \textit{Unlimited Arcade}~\textbf{\texttt{[D12]}}, a subscription-based marketplace for privacy-friendly apps, we aimed to circumvent the extreme competition present in current marketplaces. However, participants felt that Google and Apple have a sufficiently strong monopoly that such approaches will fail to gain such traction. Similarly, we had expected that \textit{Patreon and Crowdfunding}~\textbf{\texttt{[D11]}} would be positively received, as it essentially provides an additional revenue stream. However, participants did not believe that parents would see the benefits of supporting apps on such a platform and that it would run into \textcolor{black}{a} similar issue that a few apps would procure most of the funding. 

Participants were \textit{mostly negative} about approaches which tried opening a communication channel with parents. One such example is the \textit{Wisdom of the Crowds}~\textbf{\texttt{[D4]}}, a platform allowing parents to provide feedback. Participants felt that parents could not be trusted to provide \textcolor{black}{feedback as experts} and that it may introduce a policing situation, which would add even more complexities overcoming market competition. Similarly, they felt that \textit{Parental Assistance}~\textbf{\texttt{[D8]}}, which automatically generates guidance for parents about privacy risks in applications, would act as an \textcolor{black}{additional} deterrent towards potential customers, meaning developers would be punished for trying to do the right thing. 

We did not find any noticeable differences between participant subgroups in their perceptions of how our proposed ideas can benefit the developer communities, however we noticed differences in their preferences for tools they would use in their own work. For example, we found that participants who had an executive or managerial role, indicated the \textit{Requirements Matrix}~\textbf{\texttt{[D3]}} to be useful for \textcolor{black}{their work}, while developers could see themselves using SDKs, such as \textit{Plug-and-play self-control tools}~\textbf{\texttt{[D7]}}, to help speed up development. However, in their analysis of how these tools could benefit the wider development communities, they often took a viewpoint beyond their official job title, leading to varied and nuanced \textcolor{black}{perspectives}.

\begin{table}[t]

\caption{\label{table:results}  Requirements and barriers to adoption of tools and technologies to assist in privacy-friendly app development, as formulated by our study participants. }

\begin{tabular}{|p{0.45\linewidth}|p{0.5\linewidth}|}
\hline
\textbf{Needs and requirements}                                                                                                                                                                                             & \textbf{Socio(-technical) barriers    }                                                                                                                                                                    \\ \hline
Guidance for privacy-friendly app development needs to be actionable, provide international support, conflict resolution, and have clear legal implications.                                                       & Guidance should be backed by trustworthy and credible organisations which allows developers to negotiate the importance of privacy-friendly app development.                                                                                                                                     \\ \hline
SDKs should not add development overhead and should support users’ experiences.                                                                                                                                    & There is a need for users’ awareness of the importance of privacy and the normalisation of privacy-friendly app development in the development community.                                         \\ \hline
Compliance checking should happen during the development process, and should be constructive in providing actionable feedback. Compliance checking after development will undermine an app’s competitive position. & Incorporating compliance checking means additional time, effort, and money spent in app development, which means that developers should already be committed to privacy-friendly app development. \\ \hline
Need for gradual changes in the current ecosystem, which do not strongly affect users’ current habits (e.g., by having them navigate away from the mainstream marketplaces).                                       & Need for recognition of privacy-friendly development practices by end users and the need for new and more wholesome measures of success by the marketplaces.     \\            \hline                      

\end{tabular}
\end{table}

\subsection{Perceptions of guideline support}
\label{sec:guideline}

Our first set of design ideas were aimed at helping developers understand best privacy practices when designing apps for children. We focused on clarifying rules and requirements set forth by regulations (such as GDPR and COPPA), human rights organisations and public bodies (UNICEF and ICO), as well as marketplaces (Google Play and Apple App Store). We proposed three tools to address this:  \textit{Dos and Don’ts Checklist}~\textbf{\texttt{[D1]}}, \textit{Courses: Making it Official}~\textbf{\texttt{[D2]}}, and the \textit{Requirements Matrix}~\textbf{\texttt{[D3]}}. Participants reacted positively to the simplicity and non-intrusive nature of these ideas, but they also \textcolor{black}{want} sufficient technical depth to help them translate high level principles to implementation, sufficient coverage to adapt to frequent regulatory changes, and support from credible organisations to encourage adoption. 

\subsubsection{ Need for actionable guidelines with clear legal implications}
Participants confirmed findings from previous research that there is a general lack of understanding of what privacy-friendly and age-appropriate apps should look like, as well as a lack of support in addressing this gap. They found current guidelines too complex and largely inaccessible: 

\begin{quote}
   \textit{ ``I'm probably one of the few people in the world that read the entire bloody GDPR document, the whole stupid thing. I got to the end of it and it was, ‘now what?’ It's just so dense, nobody reads it."} --- P10
\end{quote}

They liked the idea of having a simplified checklist which breaks down technical rules and regulations by removing jargon and legalese. They found it important that it is accessible for a wide range of audiences involved in development, including novices and non-technical team members. The advantage of checklists is that it can provide a quick overview of what is important without having to make significant investments in reading and understanding large documents. Importantly, participants had positive experiences in working with checklists in related environments, such as quality testing in Xbox game development, which increased their confidence in having checklists for children's app development.

\begin{quote}
    \textit{`` So you know, doing Xbox game development, you have messages that you have to show in certain circumstances if you're auto saving. You absolutely have to show this icon. So having an absolute list of: `you must follow these rules as a developer', that works absolutely fine. "} --- P10
\end{quote}

However, participants also indicated that in some cases oversimplifying best practices is not practical. Not all principles are equally straightforward, with some requiring more context and technical depth. For example, participants wanted more explanations about why certain rules have to be implemented and \textit{``it would be useful in understanding where it would come from"} (P10). They wanted clarity about the origin of the principles, suggesting adding a reference or hyperlink pointing to the source document.

\begin{quote}
    \textit{`` I think saving time and giving the reference link to sort of them to go and read about it, or mock it, and it will be used. If it can be developed in a way that developers say, `I don't know about this, actually let me just go and have a look at this and take me to the right place, the right link, as opposed to just the homepage', that I think would be quite useful."} --- P2
\end{quote}

In addition, many participants expressed that it was key to minimise the friction and time between translating from principles to actual implementation, for example having exemplar implementations in the formats of mock-ups or wireframes would be tremendously helpful with \textcolor{black}{adoption of best practices}.

\begin{quote}
   \textit{``I think it'd be good to have an idea of how you would implement that, and how it might look. ... Designers particularly, when they're starting to look at this, have an idea of how it might be implemented and how we go about doing it."} --- P7
\end{quote}

Lastly, several participants described that \textcolor{black}{it is} important to have clarity on whether requirements were legally bound, nice-to-have, or enforced by marketplaces. They would like to retain choice and autonomy on what to implement, as the development of some features may be beyond the developer's budget. Moreover, when developing for clients, the development cost of every feature must be justified, which is ultimately the client's decision.

\subsubsection{Need for international regulatory support and conflict resolution} 
Because apps are often launched into an international market, a large number of participants indicated that too often they had to abide by the rules and regulations of different countries and marketplaces. However, these rules are complex, ever changing, and contain a lot of legalese, and participants indicated that clarifications in this landscape are much needed.  

\begin{quote}
    \textit{``Unity has their own terms and conditions on how you use their analytics. A lot of people don't know this, but Unity up to a particular point had their analytics for you. They were turned on no matter what you did for a long time, even with COPPA and GDPR. So, you couldn't be technically COPPA and GDPR compliant using Unity."} --- P13
\end{quote}

In addition, several of these participants also mentioned that marketplace rules change frequently and can often be unpredictable. For example, one participant explained how his app was removed from the Chinese app marketplace without warning due to a change to the rules. While this change might have been reflected in the documentation on the website, he was never notified of this. Any tool addressing legalities and marketplaces requirements could benefit from notifying developers of critical rule changes, including guidance on how to implement these changes into the app.

Lastly, some participants were worried that marketplace rules, best practices, and regulations might contradict or conflict with each other, in which case they would like guidance on how to resolve these conflicts.

\begin{quote}
    \textit{``It might be worth as well having where things might conflict. GDPR might have a regulation that conflicts with some things that Google require of apps. So that you can resolve that as best you can. "} --- P14
\end{quote}

\subsubsection{ Need for trust and credibility in courses and certifications} Participants admitted that there is a need for certification of developers' practices of designing for children. However, they were less convinced about a formal or traditional course, as they were worried it might take too much time and that companies would not be willing to pay for it:

\begin{quote}
    \textit{``The main reason why, if you don't normally get these sorts of certifications from global academic institutions like [university], you don't normally see this in an online world because it's a bit like a Wild West. There is no set rules and guidelines and certificates. There is a range of set of arbitrary skills that you can learn,  unless you do something like Udemy or whatever. But as a business owner that wouldn't carry as much weight as such."} --- P1
\end{quote}

In addition, one of the concerns participants had with adhering to privacy-friendly design principles was that they felt that `formal education' does not translate well to the wild west of the online world, where limited rules and expectations of best practices exist. Several participants thus expressed that strong credibility associated with course \textcolor{black}{certifications}, backed by governing bodies, \textcolor{black}{is} essential to offset these concerns and encourage updates by the community:

\begin{quote}
    \textit{``When it comes to safety for children, the number one thing is making sure it's from a reliable source. ... But it's more about having something that's large enough of an industry name, that everyone knows. Something like GOV.uk. I think if people had something published, that was, `this is how you do it', and `this is almost like the law, this is the best practice', that’s the best way to go forward."} --- P11
\end{quote}

\subsection{Requirements for privacy-friendly development libraries}
In order to make it easier to integrate privacy-friendly design features into apps, we proposed off-the-shelf libraries which developers can easily import and use: \textit{Plug-and-Play}~\textbf{\texttt{[D7]}}, a privacy control panel, and \textit{Parental Assistance}~\textbf{\texttt{[D8]}}, \textcolor{black}{which provides privacy information about the app} for the parents. We found that participants were largely deterred by the potential negative impact associated with integrating these tools in their app developments, for example that it may deter parents or clash with the native app branding, and therefore would prefer to wait out for a widespread adoption of these features to prevent feeling `standing out'. 

\subsubsection{ SDKs and libraries should be easy to integrate and known to be compliant}  In general, participants liked the idea of having libraries and SDKs which make privacy-friendly design easier. They indicated that developers like to rely on libraries, as it reduces development time and costs, and eliminates the need to re-implement software. Importantly, such SDKs would be guaranteed to already be compliant with any privacy-friendly feature requirements, making it a convenient and easy way to become compliant. 

However, they also pointed out that it is essential that the library is easy to use and is lightweight. One of the advantages of relying on libraries, is that it defers responsibility to an expert third party which is knowledgeable in the area of privacy-friendly development. 

\begin{quote}
    \textit{``Being able to defer a responsibility to another party, we like being able to say something is no longer our responsibility. Especially if it's something that's really technical and if it's stuff that we don't want to have to concern ourselves too much with.''} --- P10
\end{quote}

\subsubsection{ SDKs and libraries should support users' experiences} Several participants expressed concerns that the adoption of these libraries may affect users' experiences. \textcolor{black}{They feared that users' experience supported by each app could be different, which is} difficult to generalise in a third-party library. \textcolor{black}{Removing user-facing features by following the standards set in the development tools may adversely impact user experience and potentially be detrimental to user retention, sales, and other monetary aims}. Instead, they suggested implementing libraries \textcolor{black}{at} a higher level of abstraction, for example, as a skeleton framework where the underlying mechanics are functional, but the user experience can be tweaked. 

\begin{quote}
\textit{``I think it'd be good to have an idea of how you would implement that, and how it might look. I wouldn't take it as far as showing, `here's a finished thing', I'd maybe keep it as a wireframe to say, `here's some boxes of how you might do that'. So I think it'd be good to kind of get that documentation around it. "} --- P7
\end{quote}

In addition, there was a perception that privacy-friendly features which are user facing may act as a deterrent, and parents may refrain from installing the app due to its highlighted negative privacy signals, resulting in developers being penalised for doing the right thing.

\begin{quote}
    \textit{``Parents aren't thinking about it. So we're penalised for association, as the well-meaning company. And the company that hasn't got involved in parental communication about this issue, goes scot-free, because they haven't got the bandwidth to think about it."} --- P16
\end{quote}

\subsubsection{ The need for acceptance of privacy-friendly SDKs for implementing privacy-friendly features}

While participants indicated their interests in the idea of libraries and SDKs, as they can significantly reduce the threshold for engaging in privacy-friendly design practices,  participants also indicated the need for the normalisation of the use of privacy-friendly features in apps before they would consider adopting them themselves. 

Firstly, participants felt that there is little awareness amongst parents about the importance of privacy and other child-friendly features in apps. For example, they indicated that parents will not specifically seek out information presented in apps aimed at them, and in many cases, parents are not involved in the installation and onboarding process.

\begin{quote}
    \textit{``Oftentimes, the parents are not there right ... it'd be the child who opens the app for the first time, and they can't read you know, for my audience, they can't read. They are fantastic at finding out what to tap, so they will tap the screen and find a button that closes the privacy policy maybe."} --- P17
\end{quote}

As a result, they indicated that adding features aimed at parents would perhaps only confuse, or \textcolor{black}{discourage} them from using the app, meaning that costs associated with implementing such features ultimately outweigh \textcolor{black}{their} benefits. Similarly, participants were also concerned about how adding such features would situate them amongst competitors who choose not to engage in such practices, thereby perhaps not providing them with any additional competitive edge. 

We, therefore, asked participants what changes could be made that could still incentivise developers to consider integrating such privacy-friendly features. Our participants then indicated that they would be more likely to use privacy-enhancing libraries and SDKs if it had been standardised or if other developers and apps were using them as well, as this would mean that end-users would specifically seek out such behaviours in apps.

\begin{quote}
    ``\textit{It just becomes universal. `This control panel is what you have on every app that is made for children'. Because if it becomes easy for the devs to do, and then it becomes widespread, then it creates a model of expectations.}" --- P4
\end{quote}

\subsection{Concerns and needs for compliance checking mechanisms}
To help developers comply with privacy-friendly design principles, we explored design ideas for compliance checking based on (1) engaging with parents (\textit{Wisdom of the Crowds})~\textbf{\texttt{[D4]}}, (2) through an automated tool (\textit{Machine over Mind})~\textbf{\texttt{[D5]}}, and (3) outsourcing to a third party (\textit{Universal Compliance Service})~\textbf{\texttt{[D6]}}. We found that it is important for developers to receive constructive feedback from credible sources. In general, they would rather not rely on parents to be the gatekeepers of assessing privacy principles. 

\subsubsection{Feedback needs to be credible}
In general, participants liked having a tool which independently assesses apps for their privacy and age-appropriateness. Participants indicated that it was difficult to be neutral about their own app, making it more desirable to defer it to an expert third party. However, participants were concerned that a tool like \textit{Machine over Mind}~\textbf{\texttt{[D5]}} would be \textit{``nearly impossible to implement due to the variations and interpretations the system could make without the contextualization''} (P8), leading to \textit{``unreliable results and feedback''} (P3). For example, as implementations differ per app, it is difficult to assess that consent is appropriately requested. This lack of trust in the feedback and outcomes deterred some participants from using it, arguing that it might produce false positives or return erroneous feedback which would require time and labour to rectify. They were worried that even if the tool is free to use, there is a missed opportunity and time cost associated with it.

\begin{quote}
    \textit{``You know, you need to be compliant with accessibility, legislation. Which is different in the US as it is in the rest of Europe. And this is the thing that people don't get. Once you layer all these things on, you've spent a ton of your time designing for compliance, and very little of your time designing for the user. And most of these things invariably cause user attrition."} --- P13
\end{quote}

Instead, participants indicated that it is better to focus on quantitative measures, such as analysing third-party libraries and data trackers, and focus on a subset of features rather than \textcolor{black}{all aspects of privacy and} \textcolor{black}{child safety}. Similarly, participants were not keen to involve parents to review and assess apps (\textit{Wisdom of the Crowds})~\textbf{\texttt{[D4]}}. In general, they felt that parents are not the right judge to assess apps for their \textcolor{black}{privacy and age-appropriateness}. They thought parents would rather \textcolor{black}{focus} on reporting technical issues of the app, where managing these reviews can be a horrendous process.

\begin{quote}
    \textit{``Again, having worked with parents, the level of subjectivity and personal opinion is so broad. You get some people that are hyper sensitive, and even having access to the child's first name and you get some people that are happy to have all their data and stuff shared with their child on Facebook when they're six years old. "} --- P8
\end{quote}


Instead, participants preferred it if a panel of experts was appointed to assess and review apps. This way, they would be more likely to receive feedback which directly relates back to established privacy standards, rather than other subject measures of assessment. 

\subsubsection{Feedback needs to be detailed and constructive}
We also asked participants whether automated compliance checking is redundant, since Google and Apple have their own review process in place. However, a large number of participants indicated that the feedback provided by these large platforms is not sufficiently detailed, often requiring a lot of `detective work’ to figure out what is wrong. Similarly, they would not want our tools to be framed as a static safety test \textcolor{black}{that} apps are required to pass. Instead, they want constructive and detailed feedback, including suggestions for changes, which \textcolor{black}{they} are expected to be made. Participants indicated that if the tool resembles a safety test, they will try to convince or cheat the system rather than putting it to good use. For example, they would alter metadata or make superficial cosmetic changes \textcolor{black}{that ultimately do} not change the underlying functioning of the app.

\begin{quote}
    \textit{``I would be trying to convince the machine that we've done it. Not even really caring whether or not we haven't. So, well we can get a checkbox on that by doing this. ... You would end up playing the system in order to get the check boxes. So, I wouldn't care whether or not my parental controls worked."} --- P10
\end{quote}

\subsubsection{ Compliance checking needs to be effortlessly integrated with the development cycle. }
\label{res:privacy-by-design}
Participants liked the simplicity of uploading an app somewhere for assessment because it is accessible and simple to use. They compared it to security or SEO checkers \textcolor{black}{that} they have had positive experiences with before. However, participants were worried that it could produce a lot of friction and add time \textcolor{black}{to} the development process. They stated that this could happen legitimately, for example, as privacy-friendly app development might require fundamentally adding features or changing existing features, or due to errors and false positives \textcolor{black}{that} are prone to occur in such automated systems. Ultimately, if one were to use such methods, it would require them to be committed to privacy-friendly app development. \textcolor{black}{Otherwise}, they simply choose not to bring this additional work upon \textcolor{black}{themselves}.	 

\begin{quote}
    \textit{``If I really wanted to do an in-depth check, I would go to you as opposed to go to them [Google]. But the only downside with that is, you might throw up more alerts that would pass their machine, but it wouldn't pass here. So you'd be giving yourself more work to do at the end of the day. But if you were serious about what you wanted to do, you would be doing it regardless."} --- P5
\end{quote}


However, a number of participants were concerned that a test with a finished app product seemed to come at the wrong stage of the development cycle. If fundamental changes need to be made to an app, this would be costly and take a lot of time. It is unlikely that developers would engage with this. Instead, they advocated for a Privacy-by-Design approach, proposing complementing these design ideas with the \textit{Dos and Don’ts Checklist}~\textbf{\texttt{[D1]}}.

\subsubsection{ Compliance checking should not undermine an app's commercial potential}
A strong sentiment from the participants was that they did not want compliance checking of apps to negatively impact their reputation or revenue. This was particularly true with \textit{Wisdom of the Crowds}~\textbf{\texttt{[D4]}}, as they were worried that parents would suddenly become the gatekeepers of privacy-friendly apps. For example, participants felt that parents would only review apps if they have had a bad experience with \textcolor{black}{them} or that parents would focus too much on the negative aspects. 


\begin{quote}
    \textit{``I feel like parents are more likely to actively feel outraged about something as opposed to ‘I'm just going to fill this out’ right? So, I guess the thing I'm battling with is the motivation for a parent to sign up and do this."} --- P7
\end{quote}


Another common concern that participants had was that the mechanism of \textit{Wisdom of the Crowds}~\textbf{\texttt{[D4]}} could provide an opportunity for their  competitors to employ unfair methods and gain a competitive edge. For example, one participant explained how competitors left bad reviews for other apps and bought positive reviews on various marketplaces for their own apps. 

\begin{quote}
    \textit{``I started getting these negative reviews and I realised it was from the same account. Okay yeah, put a new release now and then that same account would make a negative review again. [...] What that meant was that review then became the first one people saw on the application. [...] And then I saw this person had written negative reviews about 25 other education apps. There was one particular one they hadn't written a negative review of, one that they’d written a positive review about."} --- P17
\end{quote}

\subsection{Concerns and suggestions for creating privacy-friendly apps through alternative funding mechanisms}
To explore how we could support developers seeking a balance between privacy-friendly apps and monetisation, \textcolor{black}{we proposed alternative funding schemes}, including \textit{Crowdfunding and Patreon}~\textbf{\texttt{[D11]}} , and privacy-friendly app marketplaces, \textit{Unlimited Arcade}~\textbf{\texttt{[D12]}}. \textcolor{black}{However}, participants often felt they were unlikely to succeed with these mechanisms. Instead, they proposed collaborative measures and gradual changes to leverage the current marketplaces.

\subsubsection{ Need for gradual changes in introducing alternative marketplaces}
One of the major concerns that participants had with alternative marketplaces is that \textcolor{black}{they} would directly compete with Google or Apple, because of which it is unlikely to succeed. In addition, these platforms are known to have strict rules about their apps being uploaded to other marketplaces. Because of their monopoly, participants have had \textcolor{black}{experiences} with alternative marketplaces not generating a lot of revenue. One participant explained how he uploaded his app to the Samsung marketplace. However, he earned too little money, \textcolor{black}{and} the cost of keeping track of these app versions far outweighed \textcolor{black}{their} benefits. 

\begin{quote}
    \textit{``I don't remember whether we are still on Samsung Kids. Because the network keeps changing things; I just can't keep up with updating."} --- P20
\end{quote}

For alternative platforms to be successful, participants indicated that marketing is important and that they would only consider engaging with it if it is already sufficiently popular and successful. Instead of competing with existing platforms, participants proposed leveraging their current market share and influence. For example, they suggested lobbying current platforms to buy into the scheme or to add an ethical apps category. Alternatively, they suggested creating an app library, rather than a marketplace, which redirects users to the Google or Apple marketplace.

\subsubsection{ New monetisation methods should not challenge users' existing habits }
One thing that participants liked about alternative platforms (such as Unlimited Arcade) is that \textcolor{black}{they} may raise users' awareness about the cost related to app development. Several participants indicated that there is a misconception of how much it costs to create apps. Providing more transparency on how much money developers need to raise will \textcolor{black}{}{give} parents a better understanding \textcolor{black}{of} why certain revenue sources are necessary. 
	
However, these participants were also concerned that parents are generally reluctant to pay for wholesome and privacy-friendly apps, as they are expecting everything to be free. Users' attachment \textcolor{black}{to} certain apps may lead to a market concentration around a few apps (not necessarily the best apps) in this alternative, ethical app market, while some apps will not receive funding at all.

\begin{quote}
    \textit{``And yeah, I think as a developer I'd find that mildly frustrating, that I'm now jumping through the hoops of people where I'm like, you know, I already know that parents don't quite understand the value given to them by apps."} --- P13
\end{quote}
	
Furthermore, many participants had the practical concern that parents would need to navigate away from the Apple or Google marketplace to a different environment, which could be burdensome. Instead, they suggested adding alternative ways to support crowdfunding ethical apps in the existing systems, e.g. implementing an in-app tip jar, which could avoid a complete change of users' existing habits and facilitate the objective towards developing alternative operational models for the app market.

\begin{quote}
    \textit{``I could also see something like an in-app sort of tip type thing, you know `leave developer/agency a tip'. Because then, it's sort of like: `I have used this app or my child has used this app, and it's done good'. So then they have the option to contribute for further development and maintenance."} --- P6
\end{quote}

\subsubsection{Need for new measures of success }
One of the problems participants identified with alternative marketplaces is how developers would be compensated. Currently, subscription-based platforms do this based on \textcolor{black}{gameplay} time and in-app purchases. However, nudging children to play games for longer or make purchases conflicts with children’s best interests. Instead, participants indicated that there is a need for defining new measurements for success, based on more wholesome metrics, rather than game play time. Participants were not entirely sure what this would \textcolor{black}{look} like or how this could be achieved, indicating that this could even differ per app genre. 

\begin{quote}
    ``\textit{Because a lot of the time, when it comes to the measures of success for an app, they're all around engagement and how long people have spent in the app, and how much they've bought, or all of this. So the advocacy element, [...], to not be around those things, and have more wholesome measures for success around the app.}" --- P18
\end{quote}

\subsection{Challenges for rewarding privacy-friendly apps}
To further incentivise privacy-friendly and age-appropriate app design, we explored two rewards systems: \textit{Public Pledges and Awards}~\textbf{\texttt{[D9]}}, allowing developers to take a pledge (similar to a climate pledge), and a \textit{Badge of Honour}~\textbf{\texttt{[D10]}}, \textcolor{black}{to reward} upholding certain privacy standards. Developers found that these approaches can work if there \textcolor{black}{was} sufficient community awareness of the importance of privacy, similar to climate change or organic products in the food industry.

\subsubsection{Need for recognition and adoption of privacy-friendly development practices}
In general, participants reacted favourably to the design ideas of recognising good practices. A badge or award provides positive marketing and exposure, allowing small companies to be potentially listed alongside big ones. Having a badge associated with your app, especially one awarded from a trusted or credible organisation, makes it easier to negotiate with clients and for clients to find you. Customers \textcolor{black}{benefit} from the fact that they can immediately recognise privacy-friendly apps, which allows them to make better-informed decisions.

\begin{quote}
    ``\textit{I would say any kind of way a company can have a recognised way of showing externally to their audience that they are ethical, is good. So, a badge of honour would work to show it’s certified. As long as this badge showed that was from a trusted source, and it was relatively recognised by parents.}" --- P11.
\end{quote}

However, several participants \textcolor{black}{indicated} that it was important that developers \textcolor{black}{actively apply} or seek out these badges, and that they must be assessed by an independent third party to ensure that these badges remain valuable and impactful. We asked participants whether they would be willing to pay for such a badge or its assessment, but they were concerned that this would defeat its purpose by creating a \textcolor{black}{`pay-to-win’} economy.

\begin{quote}
    \textit{``The problem that [company name] had, is that all the other reviews sites were basically based around getting money out of the app developers."} --- P17
\end{quote}

On a more practical note, a number of participants \textcolor{black}{expressed} their concerns that there are currently no negative consequences associated with apps without a badge. They compared this to the food hygiene rating practices and stated that no one will advertise a bad food hygiene rating; unless these badges are mandatory, \textcolor{black}{privacy-invasive} companies will be scarcely incentivised to cooperate.

\subsubsection{ The need to raise awareness amongst the right stakeholders}
One of the concerns that participants had \textcolor{black}{was} that it was not always clear who the target customers \textcolor{black}{are}. Participants perceived a misalignment between the customer and the buyer; children are the pursuers of games and will ultimately decide what they want to buy, however, a badge or award is unlikely to sway their decision even though their parents might take different priorities in their choice-making.

\begin{quote}
    \textit{``Kids are really driving a lot more sort of family purchasing decisions because it's technology and they're so okay with it. You're in a place, where the one, probably the person with the most knowledge is the kid. The person who needs to be protected the most is the kids, so it just creates this very interesting puzzle."} --- P16
\end{quote}

Several participants also agreed that raising awareness is not a trivial task and requires careful behaviour engineering. They made the analogy with the B Corp Certification movement (advocating for a positive impact on society) and envisioned people may slowly become more aware of the importance of privacy, suggesting that privacy-friendly apps could benefit from a similar campaign.

\begin{quote}
    \textit{``I think, as a parent, if there's somewhere that I can go or some kind of rubber stamp that I can look for, like on Twitter, they have the blue ticks. I think it's Waitrose, you can filter by B-Corp project, so you can be like: `I only want to buy from companies that are B-Corp, so I'm only going to buy from that."} --- P8
\end{quote}

Participants also proposed having different types of badges to indicate different levels or types of privacy compliance. For example, having a separate badge for data privacy or reduced game play time to support different user needs. 
\begin{quote}
    \textit{``Maybe there's different styles of badges to represent these different types of awards. Maybe they're not necessarily `this one is the best', and `this one is the worst', and there is some in between. Maybe they're just more like, different looking ones to represent what the parents might be looking for. "} --- P12
\end{quote}

\section{Discussion}

\subsection{Summary and key findings}
\begin{figure}[h]
\begin{center}
\includegraphics[width=.7\textwidth]{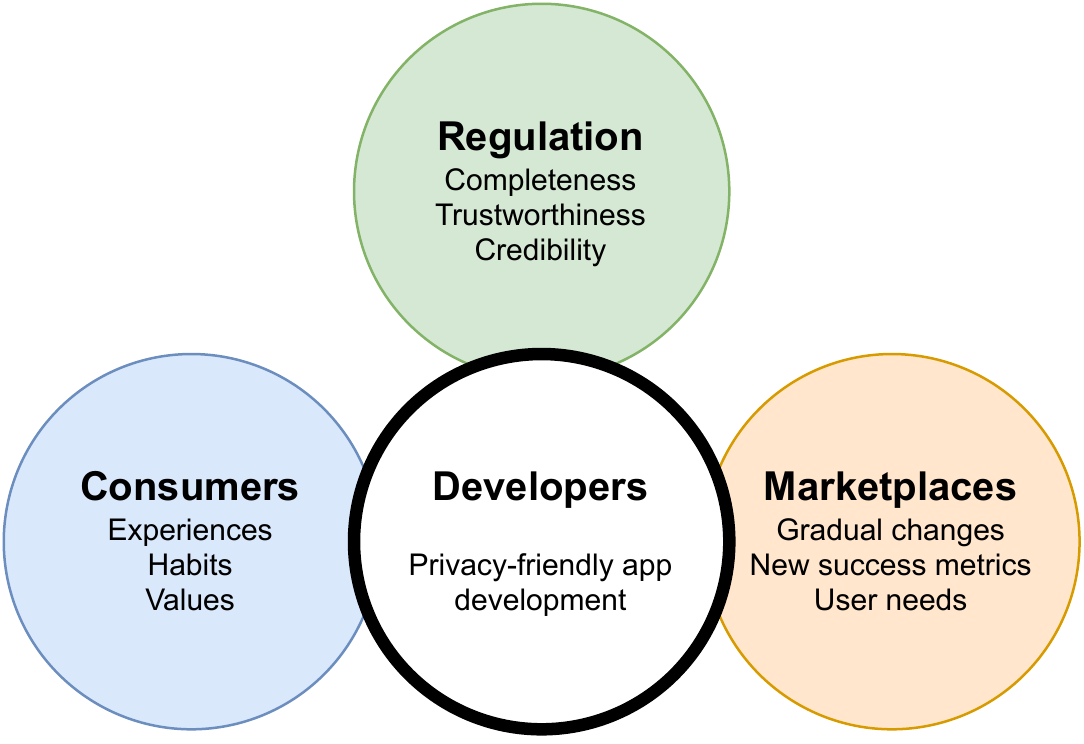}
\caption{Supporting privacy-friendly app development is a multi-stakeholder process, which requires a holistic understanding of how users' incentives intertwine with practice changes and the credibility of all stakeholders involved.}\label{fig:process}
\end{center}
\end{figure}

There is a pressing need to transform the current data-invasive approach to app development for children in order to protect their best interests and digital well-being. Using Research through Design (RtD), our study identified the critical requirements and needs of app developers \textcolor{black}{with regard to} designing privacy-friendly apps for children. These latent requirements and needs provide important guidance for designers and practitioners looking to develop support for the developer community in a specific and immediate way. However, when asked about integrating these tools into their own app development practices, participants were hesitant to fully commit to using them. They identified additional, more nuanced socio-technical barriers \textcolor{black}{which need to be} addressed before they would consider adopting privacy-friendly development practices. Our participants perceived these barriers by considering the values and roles of multiple stakeholders, including parents, children, regulatory organisations, and current and future platform providers. Through this, we identified a critical dimension for future development of support for app development: \textit{the importance of multi-stakeholder engagement}. Addressing developers' barriers and incentives, developing trust in design guidelines to support them in navigating and bringing change to the monopolising market space, and other future design explorations all require critical engagement beyond the developer community and with multiple stakeholders. This reflects the fact that successful privacy-friendly app development must be situated among multiple paradigm shifts and behaviour engineering (see Figure~\ref{fig:process}). In the following sections, we delve deeper into the socio-technical requirements and barriers.

\subsection{Addressing the need for credibility and normalisation of privacy-friendly app development}
\label{discussion:guidelines}
\textcolor{black}{Developers often perceive a trade-off between implementing privacy features and adhering to guidelines versus financial benefits.}. In some cases, these costs were explicit. For example, as was the case with our automated compliance testing suite, \textit{Machine over Mind}~\textbf{\texttt{[D5]}}, making changes retroactively would take a lot of time and effort. In other cases, these costs were implicit or hidden, for example through friction encountered during development, even when integrating or using existing SDKs. As a result, participants found it important that the costs associated with creating privacy-friendly apps are offset with certain benefits. 


Firstly, they required that their efforts are recognised by consumers. They proposed that guidelines and tools be backed by influential organisations as a way to communicate trustworthiness and credibility. There has been an increasing number of efforts to establish guidelines and tools aimed at developers. For example, UNICEF published the policy guidance on AI for children \cite{dignum2020policy}, detailing how AI systems should uphold children’s rights and \textcolor{black}{offering} a set of practical implementation tools with it, such as a development canvas. UNICEF also offers a similar toolkit for privacy \cite{nyst_gorostiaga_geary_2018}, which aims to clarify the GDPR and which contains a questionnaire-based checklist, covering the full spectrum of privacy considerations for children \textcolor{black}{that} developers must make. Similarly, the ICO in the UK provides tools to help developers understand how children's data is mapped in their applications \cite{ico_data_maps}, as a way to help implementing the Age-Appropriate Design Code and be more intentional with data processing. Despite these efforts, studies have shown that developers find it difficult to keep track of new guideline developments and integrate them into a common compliance framework \cite{ekambaranathan2021money}. \textcolor{black}{Developers require practical guidelines that resolve conflicts between frameworks, as shown in studies assessing Privacy Engineering Methods \cite{senarath2019will}. However, our study suggests that practical elements should also consider commercial aspects, as developers negotiate privacy concerns with consumers. Current guidelines should involve additional stakeholders, such as developers, consumers, and domain-specific experts from cybersecurity or child development \cite{wisniewski2017parental,nouwen2015value}.} This aligns with various technical guideline development approaches \cite{assal2019think}, which is supported \textcolor{black}{by} a rich body of HCI research \textcolor{black}{on} how software development can benefit \textcolor{black}{from} integrating privacy principles into the development process \cite{tahaei2021privacy, baldassarre2020integrating}. Participatory design of guidelines and tools would ensure that they are formatted according to developers' needs, \textcolor{black}{so} that the time and effort to use them \textcolor{black}{are} minimised. 


\textcolor{black}{Secondly, developers desire normalisation of privacy-friendly design features in apps, and suggest standardisation efforts to establish common expectations among competitors and consumers. Credible organisations, such as the IEEE Standards Association \cite{ieeesa} and the UK ICO \cite{ico_cert}, can enforce compliance and validation through certification programmes. This enables independent bodies to carry out enforcement more efficiently.} However, certifications for privacy compliance are mainly focused on legal requirements, rather than positioning efforts in the marketplace where developers and consumers can benefit. To address this, both consumers and developers should be considered as stakeholders when developing standards. Other means to normalise privacy include corporate-level efforts and placing privacy champions in software teams \cite{iwaya2022privacy,tahaei2021privacy}. However, these efforts may not include independent developers or small teams. Therefore, a more holistic approach is needed to put systems in place that benefit all stakeholders, such as raising awareness among parents, children, and developers through multi-disciplinary approaches, in which the HCI community has an important role to play in this.

\subsection{Users' incentives, behaviours and new practices}
Some of our design approaches, such as the \textit{Badge of Honour}~\textbf{\texttt{[D10]}}, were motivated by existing research related to privacy nutrition labels \cite{kelley2009nutrition,van2017better}, and were specifically aimed at helping developers raise awareness amongst their end users about the improved privacy aspects of their apps. While existing research has shown how the effectiveness of privacy nutrition labels is correlated with users' level of privacy awareness~\cite{kelley2009nutrition,van2017better} and when they were presented to the users~\cite{almuhimedi2015your,acquisti2017nudges}, our results showed that our participants were more concerned whether users were prepared to make any actual changes of their app purchase habits (i.e. incentive changes) and the credibility of such `badges of honour'.

\textcolor{black}{Participants compared our design proposals to movements in similar consumer industries, such as Fairtrade \cite{fairtrade} and B-Corp \cite{bcorp}. However, these approaches require a multi-stakeholder audience to change their behaviour towards privacy, which is currently not a priority for parents when choosing apps for their children \cite{broekman2016parental}. Safety and privacy are often considered separate concerns, but as our study participants indicated, all stakeholders must understand the consequences of a loss of privacy \cite{longfield2018knows}. Normalising discussions about privacy is crucial, for example, by including it in school curricula or conversations between parents and children \cite{kumar2019privacy}.}


To overcome the barriers for the recognition of new, privacy-friendly development practices, developers in our study indicated the importance of a two-step approach: (1) establishing a market standard for \textit{a privacy-friendly app for children}, and (2) promoting the recognition of this standard by the end users (such as parents). In this way, their additional effort would lead to visible recognition amongst their competitors and users and ensure a possible competitive edge. We see a resemblance from this call for socio-technical changes to previous privacy research~\cite{kelley2009nutrition,almuhimedi2015your} which have found that increasing the transparency of personal data being collected by websites or smartphone devices for the users would not necessarily effectively impact users' privacy behaviours unless their awareness about this transparency and its implications were securely established. However, facilitating this user awareness and community change in the app ecosystem poses some unique challenges. 

\textcolor{black}{Rather than a `privacy paradox' \cite{norberg2007privacy}, research has identified a 'value paradox' among parents in supporting their children's online activities. Despite valuing their education and media interaction, they often underestimate the cost of good apps and are reluctant to pay for any \cite{livingstone2018parenting}.} This can be partially attributed to the \textcolor{black}{freemium model}, which has dominated and influenced consumer behaviour for a substantial period of time \cite{liu2014effects}. \textcolor{black}{The lack of backing from credible organisations may hinder the importance of privacy practices for developers. To support developers in overcoming this barrier, it is important to encourage them to reconsider the factors that influence their development decisions. Future investigations should focus on integrating users' awareness and recognition to boost developers' motivations.}


\subsection{Overcoming systemic and structural barriers} 
 
Major industry stakeholders, such as Apple and Google, are poorly incentivised to limit data collection and analytics, as it forms a major source of their revenue \cite{acquisti2016economics}. Both Apple and Google do not react kindly to any initiatives trying to undermine their terms and conditions, as was demonstrated by Epic Games v. Apple court case \cite{smizer2021epic}. \textcolor{black}{Epic Games cut Apple's 30\% commission on in-app purchases, resulting in Apple blocking them from their marketplace. Smaller studios cannot afford legal action like Epic Games. Apple has taken an opposing stance to Google's data-driven approach by introducing initiatives such as privacy labels and requiring apps to ask permission for third-party tracking services \cite{apple_att,apple_labels}. Major industry players are aware of the growing concerns about privacy and are shifting away from data-driven monetisation methods. Apple has an advantage, as they rely less on advertising revenue compared to Google or Facebook. Other marketplace providers must show similar support to facilitate a shift to privacy-friendly apps.}


While our participants welcome the vision of creating alternative marketplaces, they raised the concerns regarding the current measurement of app success, which is commonly based on gameplay time. Without a fundamental rethinking and promotion of alternative ways of evaluating apps' values and contributions, the community could hardly foresee a successful paradigm shift and associated practice changes. The major platforms have started introducing subscription based services, in which games can be accessed without in-app purchases and advertising \cite{apple_arcade, google_play_pass}. While these platforms are not transparent about their payment structure, interviews with app developers have revealed that royalties are directly still tied to gameplay time, incentivising app developers to build in addictive feedback loops and other features to maximise gameplay time \cite{Kim_2019}. Additionally, developers expressed concerns that these payment structures only benefit big development studios, leaving apps with a small user base earning very little.

Currently there are no platforms in existence where alternative payment structures are considered, especially since Google and Apple have a monopoly in this field. \textcolor{black}{This suggests} the need for a paradigm shift in our models for data governance. One example of this is the Web 3.0 concept \cite{ragnedda2019blockchain}, which is aimed to help facilitate decentralised data architectures and restore data autonomy back to users. For example, decentralised marketplaces allow buyers and sellers to exchange `goods' without the need of central service providers, which in the mobile ecosystem would allow the community to eliminate Apple and Google. It would also give developers the autonomy to establish how they wish to monetise their services, rather than being forced to rely on in-app purchases and targeted advertising. 

\textcolor{black}{However, introducing a paradigm shift in app development may require structures to overcome the dominance of leading app markets \cite{robertson_2021}. For example, participants suggested that alternative markets, like F-Droid, would be critical for redesigning the data governance structure. At the same time, this shift raises new research challenges in facilitating the creation and uptake of this new data governance paradigm, to ensure users regain their data rights.}


\subsection{Limitations and future work}
The methodology we used had several limitations. Firstly, we focused on the UK app market, because of new regulatory initiatives introduced in late 2020 for organisations developing digital products and services aimed at children. However, given the international nature of app marketplaces, we encourage future studies to incorporate stakeholders from other geographic areas as well. 

\textcolor{black}{Secondly, we focused on mobile apps, rather than a broader range of products and services for children.} We have not specifically asked participants whether they perceived any difference for developing for children of different ages. Our focus has been on the data protection and privacy aspects of children's apps instead of the more nuanced support for different developmental needs of children. 

Thirdly, in our analysis of the interviews, we did not make a distinction between subgroups of participants (e.g., work experience or role). We acknowledge that there were several variations in job roles, however all participants have had experience in creating and publishing apps for children. In our study, they often used their app creators' lens to assess and evaluate our design ideas. However, they did indicate individual preferences of design ideas as they would benefit different roles in different ways. 

Fourthly, we used a sample size of 20 developers, primarily from the UK, meaning that our results may not generalise to a wider international audience of developers, where regulatory requirements such as the GDPR, COPPA, and the AADC are less prioritised. In our case, as our focus was on the UK market, our in-depth interviews with a limited sample size provided rich data. Future studies may consider using methods to reach larger audiences, such as surveys, and even venture beyond the UK/EU developer communities. 

Lastly, our findings were closely derived from the 12 design ideas that we generated in the ideation workshop. While the ideas were sufficiently diverse, it would have been interesting to include developers (and other stakeholders) as co-speculators in the design process itself. However, the community for developing children's apps specifically is rather limited in the UK and the consequences of the COVID-19 pandemic made it difficult to gather such a unique set of stakeholders in a room together. However, for future studies, it would be interesting to co-design solutions directly with stakeholders, and even prototype some of the design ideas to evaluate how it would fare in practice. 
\section{Conclusion}
There is an increasing demand from consumers and regulatory bodies for apps and services aimed at children to make data protection and privacy a core element of their design. However, mobile app developers are not sufficiently empowered to understand and integrate privacy preserving design features in their products due to conflicts of interest between marketplace providers and other stakeholders. In this study, we examined requirements to enable developers to address these challenges by eliciting reactions to 12 speculative design concepts we created. Our findings show that developers are in need of tools to help them design privacy-friendly apps. However, they also formulated critical socio-technical barriers preventing adoption of these tools. Overcoming these barriers requires a substantial effort from the HCI design community, relying on multi-stakeholder multi-disciplinary initiatives. We examined in depth how developers envision the future of privacy-friendly app development to look like, and how we can overcome some of the socio-technical barriers they formulated.


\end{document}